\documentclass[twocolumn,conference,english]{IEEEtran}
\usepackage[T1]{fontenc}
\usepackage[utf8]{luainputenc}
\usepackage{babel}
\usepackage{verbatim}
\usepackage{amsthm}
\usepackage{amsmath}
\usepackage{amssymb}

\usepackage[unicode=true,
 bookmarks=true,bookmarksnumbered=true,bookmarksopen=true,bookmarksopenlevel=1,
 breaklinks=false,pdfborder={0 0 0},backref=false,colorlinks=false]
 {hyperref}
\hypersetup{pdftitle={Your Title},
 pdfauthor={Your Name},
 pdfpagelayout=OneColumn,pdfnewwindow=true,pdfstartview=XYZ,plainpages=false}

\makeatletter
\theoremstyle{plain}
\newtheorem{thm}{\protect\theoremname}
\theoremstyle{plain}
\newtheorem{lem}[thm]{\protect\lemmaname}
\theoremstyle{plain}
\newtheorem{cor}[thm]{\protect\corollaryname}

\usepackage{babel}
\ifCLASSOPTIONcompsoc
\else
\fi
\DeclareMathOperator*{\argmax}{arg\,max}

\providecommand{\corollaryname}{Corollary}
\providecommand{\lemmaname}{Lemma}
\providecommand{\theoremname}{Theorem}

\makeatother

\providecommand{\corollaryname}{Corollary}
\providecommand{\lemmaname}{Lemma}
\providecommand{\theoremname}{Theorem}

\begin{document}

\title{New Algebraic Soft Decision Decoding Algorithm for Reed-Solomon Codes}

\author{\IEEEauthorblockN{Yuan Zhu}\\
\vspace{-0.5cm}
\IEEEauthorblockA{Email: 41656955@qq.com}
\and
\IEEEauthorblockN{Siyun Tang} \\
\vspace{-0.5cm}
\IEEEauthorblockA{GuangDong Polytechnic Normal University, Guangzhou, China\\
Email:tangsiy@mail2.sysu.edu.cn}
\vspace{-5.5cm}
}
\vspace{-1.5cm}
\IEEEspecialpapernotice{\vspace{-1cm}
}
\maketitle
\vspace{-1.5cm}
\IEEEspecialpapernotice{\vspace{-1cm}
}
\vspace{-0.5cm}
\begin{abstract}In this paper, we present a new algebraic soft-decision decoding algorithm for
Reed-Solomon codes. It is based on rational interpolation
and the interpolation points are constructed by Berlekamp-Messay algorithm.
Unlike the traditional K{\"o}tter-Vardy algorithm, new algorithm needs
two smaller multiplicity matrixes for interpolation, due to the new
factorization algorithms for re-constructing codewords.

\end{abstract}

\vspace{-0.3cm}

\section{Introduction}

\IEEEPARstart{R}{eed}-Solomon code, or RS code, is widely used
in communication and digital storage. Its definition is based on the
algebraic theory in finite field \cite{RS_code}. So most of its decoding
algorithms are based on related algebraic theory. Since Peterson's
invention\cite{PetersonDecoding} and other work \cite{BM,eucidian_algorithm},
the hard-decision bounded-decoding algorithms for RS code are widely
used in practice. Modern RS code decoding algorithms, including list-decoding
algorithm \cite{GS1999} and algebraic soft-decision decoding algorithm
\cite{KV2003}, enable better performance. These algorithms are based
on polynomial interpolation with special multiplicity. A new list-decoding
algorithm \cite{Wu2008} for RS code based on rational interpolation
has been introduced in 2008. It needs smaller interpolation multiplicity
for the same list-decoding radius. In that paper, an open problem
is raised: can we generalize the new algorithm for algebraic soft-decision
decoding? In this paper, we give an algorithm to answer this question.%
{} %

\vspace{-0.2cm}
\section{Previous Work}
\vspace{-0.1cm}

The traditional error correction algorithm of $(n,k)$ RS code in
$GF$($q$) is to get an error locator polynomial.
Suppose $\alpha$ is a primitive element of $GF$($q$) and $n=q-1$. For a hard-decision vector of the receiving symbols, if
it is different from the sending codeword at error locations set $I\subseteq\{1,2,\ldots,n\}$,
then we have the error locator polynomial $\Lambda(x)=\prod_{i\in I}(1-\alpha^{i}x)$ for the codeword.
$\Lambda(x)$(or $\Lambda$ for short, like other function in the following content) is also a connection polynomial for the syndromes sequence
$(S_{0},S_{1},\ldots,S_{n-k-1})$ while the syndromes sequence is
the first $n-k$ part of the Fourier transformation of all possible
error patterns.

The well known BM algorithm \cite{BM} can generate the error location
polynomial $\Lambda$, if the weight of error pattern is not beyond
$\left\lfloor(n-k)/2\right\rfloor$. It always outputs two co-prime polynomial: one is
$\lambda$, the connection polynomial of syndromes sequence with minimal
degree; the other is $\delta$, the correction polynomial. What's
more, $\deg(\lambda)+\deg(x\delta)\leq L_{\lambda} + L_{x\delta}=n-k+1=d$, $L_{\lambda}$ and  $L_{x\delta}$ is the length of linear feedback shift register described by $\lambda$ and $x\delta$. From
Wu's result \cite[Lemma 3]{Wu2008}, every possible $\Lambda$ can be
expressed as $\Lambda=\lambda u+x\delta v$ , $\deg(u)\leq \deg(\Lambda)-L_{\lambda},\deg(v)\leq \deg(\Lambda)-L_{x\delta}$,
$u$ and $v$ are co-prime polynomials and $u(0)=1$. By the fact
that the evaluation of $\lambda u+x\delta v$ equals to zero at and
only at the error locations, Wu's algorithm uses a bivariate polynomial
$Q(x,y)$ with minimal $(1,L_{\lambda}-L_{x\delta})$- weighted
degree to interpolate the $n$ points $\{(\alpha^{-i},-\frac{\lambda(\alpha^{-i})}{\alpha^{-i}\delta(\alpha^{-i})})|i=1,2,\ldots,n\}$
with special multiplicity (if $\delta(\alpha^{-i})=0$, the $i$th
interpolation point is ($\alpha^{-i},\infty$)). $uy-v$ will be a
factor of $Q(x,y)$ if the weight of the error pattern is not beyond
the list-decoding radius of Wu's algorithm. As $u(0)=1$, when $Q(x,y)$
has factor $uy-v$, we can use Roth-Ruckenstein factorization algorithm
\cite{RR_factorization} to get first $\deg(u)+\deg(v)$ terms of
Maclaurin series of $v/u$ and Pad{\'e} approximation to get
$v/u$ \cite{Wu2008}.

Some relative algorithms generate the interpolation points not from
BM algorithm, but extended Euclidean algorithm \cite{Kuijper,Trifonov_Wu}.
In fact, they have the same performance for the same decoding \textbf{radius} as Wu's algorithm
in the worst case, either measured by the interpolation multiplicity
or the list decoding size. It is usually ignored when the maximal $y$-degree of the interpolation result is greater than the multiplicity, the decoding radius is limited by the possible $\left\lfloor(n-k)/2 + 1\right\rfloor$ weight error pattern (from \cite[Lemma 3]{Wu2008}).

Something has to be mentioned before further discussion. First, as
there may exist interpolation points at infinity, we can not use the
traditional module based bivariate interpolation algorithm (including
K{\"o}tter algorithm\cite{Koetter_interpolation} and Lee-O'Sullivan
algorithm\cite{lee_Osullivian}) to do the interpolation because the
legal interpolation result set with finite $y$ degree limit is
not a module. Second, if we can not guarantee $u(0)\neq0$, we can
not get Maclaurin series of $v/u$, which is not mentioned
in some previous work \cite{Kuijper,Trifonov_Wu}.


\vspace{-0.1cm}

\section{Improved Rational Interpolation Algorithm}

We first construct new rational interpolation points without potential
infinity. If $L_{\lambda}\leq L_{x\delta}$, we denote $a=\lambda$
and $b=x\delta$; or else $a=x\delta$ and $b=\lambda$. Without loss
of generality, suppose $\Lambda=au+bv$. Let $f(x)=b-\theta a$, $\theta\in GF(q)$
and we define $f(x)=a$ if $\theta=\infty$.
\vspace{-0.1cm}
\begin{lem}
There exist at least two different value $\theta_{1},\theta_{2}\in(GF(q)\bigcup\{\infty\})$
to be $\theta$, so that $f(x)=0$ does not have any root in $\{\alpha^{-1},\alpha^{-2},\ldots,\alpha^{-n}\}.$\end{lem}
\begin{IEEEproof}
As $a$ and $b$ are co-prime, if $f(x_{i})=0$, then $\theta=\frac{b(x_{i})}{a(x_{i})}$.
Obviously, $b/a$ can have at most $n$ different evaluation results
(including $\infty$) from $x=\alpha^{-1},\alpha^{-2},\ldots,\alpha^{-n}$.
As $GF(q)\bigcup\{\infty\}$ has two more elements than probable evaluation
results, so there must exist at least two different values $\theta_{1},\theta_{2}\in(GF(q)\bigcup\{\infty\})$
to be $\theta$, so that $f(x)=0$ does not have any root in $\{\alpha^{-1},\alpha^{-2},\ldots,\alpha^{-n}\}.$
\end{IEEEproof}
We can select one value from $\theta_{1}$ and $\theta_{2}$ as $\theta$.
If $\theta\neq\infty$, $\Lambda=a(u+\theta v)+(b-\theta a)v$. Obviously, for $i\in I$, 
\[
\deg(u+\theta v)\leq\max(\deg(u),\deg(v))\leq\deg(\Lambda)-L_{a}
\]
\[
\frac{v(\alpha^{-i})}{u(\alpha^{-i})+\theta v(\alpha^{-i})}=-\frac{a(\alpha^{-i})}{b(\alpha^{-i})-\theta a(\alpha^{-i})}\neq\infty
\]

If $\theta=\infty$,
\[
\frac{u(\alpha^{-i})}{v(\alpha^{-i})}=-\frac{b(\alpha^{-i})}{a(\alpha^{-i})}\neq\infty
\]

Let's define $g(x)$ and $h(x)$:
\[
g(x)=\begin{cases}
\frac{v}{u+\theta v} & ,\theta\neq\infty\\
\frac{u}{v} & ,\theta=\infty
\end{cases},
h(x)=\begin{cases}
-\frac{a}{b-\theta a} & ,\theta\neq\infty\\
-\frac{b}{a} & ,\theta=\infty
\end{cases}
\]

When a bivariate polynomial $Q(x,y)$ interpolates the $n$
points $\{(\alpha^{-1},h(\alpha^{-1})),(\alpha^{-2},h(\alpha^{-2})),\ldots,$
$(\alpha^{-n},h(\alpha^{-n}))\}$ at special multiplicity, as $\deg(u+\theta v)$ has same upper bound as $\deg(u)$,
we
can get the similar result like Wu's algorithm. We use $\deg_{w_{x},w_{y}}(Q)$
to denote the $(w_{x},w_{y})$-weighted degree of $Q(x,y)$, $D_{Y}(Q)=\deg_{0,1}(Q)$.
\begin{thm}
For $\theta\neq\infty$ and $Q(x,y)$ that interpolates
$n$ points $\{(\alpha^{-1},h(\alpha^{-1})),(\alpha^{-2},h(\alpha^{-2})),\ldots,(\alpha^{-n},h(\alpha^{-n}))\}$
with multiplicity $r$, If $\deg_{1,L_{a}-L_{b}}(Q)+(\deg(\Lambda)-L_{a})D_{Y}(Q)<r\deg(\Lambda)$,
then $Q(x,g(x))=0$.\end{thm}
\begin{IEEEproof}
If $\theta\neq\infty$, suppose $\deg(u+\theta v)=\zeta$, then
the maximum degree of numerator of $Q(x,g(x))$ is $\deg_{1,\deg(v)-\zeta}(Q)+\zeta*D_{Y}(Q)$,
which is monotone increasing by $\zeta$. With $\deg(v)\leq \deg(\Lambda)-L_{b}$, $\zeta \leq \deg(\Lambda)-L_{a}$, the maximum
degree of numerator of $Q(x,g(x))$ is $\deg_{1,L_{a}-L_{b}}(Q)+(\deg(\Lambda)-L_{a})D_{Y}(Q)$.
For $i\in I$, $h(\alpha^{-i})=g(\alpha^{-i})$, so the numerator
of $Q(x,g(x))$ passes $x=\alpha^{-i}$ with multiplicity $r$. Once
the degree of a polynomial is less than the sum of its zero points'
multiplicity, it must be the zero polynomial.\end{IEEEproof}
\begin{cor}
For $\theta=\infty$ and $Q(x,y)$ that interpolates $n$
points $\{(\alpha^{-1},h(\alpha^{-1})),(\alpha^{-2},h(\alpha^{-2})),\ldots,(\alpha^{-n},h(\alpha^{-n}))\}$
with multiplicity $r$. If $\deg_{1,L_{b}-L_{a}}(Q)+(\deg(\Lambda)-L_{b})D_{Y}(Q)<r\deg(\Lambda)$,
then $Q(x,g(x))=0$.
\end{cor}
Both Theorem 2 and Corollary 3 does not need $Q(x,y)$ to pass interpolation
points at infinity. So all the module based bivariate polynomial interpolation
algorithm can be used and we do not have any performance loss from Wu's result.

We may notice that as $u$ and $v$ are co-prime, so are $u+\theta v$
and $v$ when $\theta\neq\infty$. As rational polynomial set is a
field, when $Q(x,g(x))=0$, if $\theta\neq\infty$, then $Q(x,y)$
has the factor $(u+\theta v)y-v$; or else it has factor $vy-u$.
However, we can not directly use Wu's factorization algorithm because
we can not guarantee the denominator of $g(x)$ does not has factor
$x$, which is required to get the Maclaurin series.
In another word, $g(x)$ may have pole in any element in $GF$($q$)
so we are not sure to be able to get its Taylor series at any point.
But we can do some transformation to avoid the problem. If $a=\lambda$
and $\theta\neq\infty$, let $\hat{Q}(x,y)=Q(x,y^{-1})y^{D_{Y}(Q)}$,
then we can do factorization from $\hat{Q}(x,y^{-1}+\theta)y^{D_{Y}(\hat{Q})}$
to get factor $uy-v$ (if $\theta=0$, we can directly factorize $Q(x,y)$);
if $b=\lambda$ and $\theta\neq\infty$, then we can do factorization
from $Q(x,y^{-1})y^{D_{Y}}$ to get factor $vy-(u+\theta v)$; if
$a=\lambda$ and $\theta=\infty$, then we can do factorization from
$Q(x,y^{-1})y^{D_{Y}}$ to get factor $uy-v$; if $b=\lambda$ and
$\theta=\infty$, then we can directly factorize $Q(x,y)$ to get
factor $vy-u$.

We need to pay attention that there are at least two choices of $\theta$.
For list-decoding in hard decision, we can choose any one of them. However,
the other one is not useless. In the following soft-decision decoding
algorithm, we need both of them to do interpolation twice.

\section{Algebraic Soft Decision Decoding Algorithm}

\subsection{Construction of Interpolation Points }

Let's denote the Fourier transformation of the error pattern as polynomial
$E$. By the property of Fourier transformation and reverse, the value
of error pattern at location $\alpha^{-i}$ is $-E(\alpha^{-i})$.
So $\Lambda E$ has the factor $\prod_{i=1}^{n}(1-\alpha^{i}x)=1-x^{n}$.
Then we can define the error evaluator polynomial $\Omega$ that $\Lambda E=\Omega(1-x^{n})$.
Because the degree of $E$ is less than $n$, the degree of $\Omega$
is less than $\Lambda$. For any natural number $\gamma\leq n$,
\vspace{-0.3cm}

\begin{equation}
\Lambda(E\mod x^{\gamma})\equiv\Omega\mod x^{\gamma}\label{eq:generalized key-equation}
\end{equation}
\vspace{-0.6cm}

If the $n-k$ syndromes sequence $(S_{0},S_{1},\ldots,S_{n-k-1})$
can be viewed as a polynomial $S=\sum_{i=0}^{n-k-1}S_{i}x^{i}$, then $S=E\mod x^{n-k}$. And we
have the traditional key-equation

\vspace{-0.3cm}

\[
\Lambda S\equiv\Omega\mod x^{n-k}
\]

\vspace{-0.3cm}

An alternative algorithm to solve the key-equation is called Berlekamp
algorithm \cite{Berlykampbook,blahut2008algebraic}. It computes iteratively
not only $\Lambda$ but also $\Omega$. While BM algorithm begins
with $\lambda=1$, $\delta=1$, Berlekamp algorithm begins with $\lambda=1$,
$\delta=1$, $\omega=0$, $\kappa=x^{-1}$. Berlekamp algorithm shares
the same update rule of $(\lambda,\delta)$ as it in BM algorithm,
and the update of $(\omega,\kappa)$ also follows that rule. For example,
if $(\lambda,\delta)$ is updated as $(\lambda,\delta)\leftarrow(\lambda+\Delta\delta,\lambda)$ in an iteration, then $(\omega,\kappa)$
is updated as $(\omega,\kappa)\leftarrow(\omega+\Delta\kappa,\omega)$.
We may notice $\Omega=\Lambda(E(1+x^{n}+x^{2n}+\cdots))$. It's proved
that once we know the syndromes sequence from coefficient of $E(1+x^{n}+x^{2n}+\cdots)\mod x^{2\deg(\Lambda)}$,
then $\Omega=\omega$ and $\Lambda=\lambda$ after $2\deg(\Lambda)$
iterations of Berlekamp algorithm \cite{Berlykampbook}. If $a=\lambda$
and $b=x\delta$, we denote $c=\omega$ and $d=x\kappa$; or else
$c=x\kappa$ and $d=\omega$. By the fact that $\Lambda$ and $\Omega$
can be generated by the same transformation from $(\lambda,\delta)$
and $(\omega,\kappa)$ during the additional $2\deg(\Lambda)-(n-k)$
iterations when $2\deg(\Lambda)-(n-k)>0$, obviously we have the following
conclusion.
\vspace{-0.2cm}
\begin{lem}
If %
{} $\Lambda=au+bv$ , then $\Omega=cu+dv$.
\end{lem}
For $i\in I$, $-E(\alpha^{-i})=-\frac{(\Omega(1-x^{n}))^{\prime}}{\Lambda^{\prime}}\mid_{x=\alpha^{-i}}$
by L{\^o}pital's rule, or
\vspace{-0.5cm}
\begin{eqnarray*}
-E(\alpha^{-i}) & = & -\frac{\Omega(\alpha^{-i})}{\alpha^{-i}\Lambda^{\prime}(\alpha^{-i})}\\
 & = & -\frac{cu+dv}{x(au+bv)^{\prime}}\mid_{x=\alpha^{-i}}
\end{eqnarray*}

We denote error value $e_{i}=-E(\alpha^{-i})$. When $\theta\neq\infty$,
as $h(\alpha^{-i})=\frac{-a(\alpha^{-i})}{b(\alpha^{-i})-\theta a(\alpha^{-i})}=\frac{v(\alpha^{-i})}{u(\alpha^{-i})+\theta v(\alpha^{-i})}=g(\alpha^{-i})$,
so
\begin{alignat*}{1}
(-e_{i}\alpha^{-i})^{-1} & =\frac{(a(u+\theta v)+(b-\theta a)v)^{\prime}}{c(u+\theta v)+(d-\theta c)v}\mid_{x=\alpha^{-i}}\\
 & =\frac{(-h(b-\theta a)(u+\theta v))^{\prime}}{c(u+\theta v)+(d-\theta c)v}\mid_{x=\alpha^{-i}}\\
 & +\frac{(g(b-\theta a)(u+\theta v))^{\prime}}{c(u+\theta v)+(d-\theta c)v}\mid_{x=\alpha^{-i}}\\
 & =\frac{(b-\theta a)(g^{\prime}-h^{\prime})}{c+(d-\theta c)h}\mid_{x=\alpha^{-i}}
\end{alignat*}

When $\theta=\infty$, $h(\alpha^{-i})=-\frac{b(\alpha^{-i})}{a(\alpha^{-i})}=\frac{u(\alpha^{-i})}{v(\alpha^{-i})}=g(\alpha^{-i})$,
\begin{eqnarray*}
(-e_{i}\alpha^{-i})^{-1} & = & \frac{(au+bv)^{\prime}}{cu+dv}\mid_{x=\alpha^{-i}}\\
 & = & \frac{(gav-hav)^{\prime}}{cu+dv}\mid_{x=\alpha^{-i}}\\
 & = & \frac{a(g^{\prime}-h^{\prime})}{d+ch}\mid_{x=\alpha^{-i}}
\end{eqnarray*}

Let's define $\phi(x)$:
\[
\phi(x)=\begin{cases}
\frac{c+(d-\theta c)h}{b-\theta a}=\frac{-ad+bc}{(b-\theta a)^{2}} & ,\theta\neq\infty\\
\frac{(d+ch)}{a}=\frac{ad-bc}{a^{2}} & ,\theta=\infty
\end{cases}
\]
For $i\in I$, we can evaluate the error value if we get $g^{\prime}(x)$ :
\begin{equation}
e_{i}=-\frac{\phi(\alpha^{-i})}{\alpha^{-i}(g^{\prime}(\alpha^{-i})-h^{\prime}(\alpha^{-i}))}\label{eq:error value}
\end{equation}
We use $p(x,e)$ to construct the interpolation points for $g^{\prime}(x)$
\begin{align*}
p(x,e) & =-(ex)^{-1}\phi(x)+h^{\prime}(x), & \quad e\neq0,x\neq0
\end{align*}
We can prove $\phi(\alpha^{-i})$ will never be zero for any $i$.
The numerator of $\phi(x)$ is $ad-bc$. If $\frac{b(\alpha^{-i})}{a(\alpha^{-i})}\neq\frac{d(\alpha^{-i})}{c(\alpha^{-i})}$,
then $\frac{b(\alpha^{-i})+\Delta a(\alpha^{-i})}{a(\alpha^{-i})}\neq\frac{d(\alpha^{-i})+\Delta c(\alpha^{-i})}{c(\alpha^{-i})}$.
Before the first iteration, $a=1$, $b=x$, $c=0$, $d=1$, $\frac{b(\alpha^{-i})}{a(\alpha^{-i})}\neq\frac{d(\alpha^{-i})}{c(\alpha^{-i})}$
for $i=1,2,\cdots,n$. So by induction in Berlekamp algorithm, $\phi(\alpha^{-i})$
will never be zero for any $i$. In another word, for different value
of $e$ and fixed value of $x$, the evaluation of $p(x,e)$ is different.

If we can have a multiplicity assignment function $M(x,e)$, which
maps from $F^{2}$ to $Z^{+}\cup\{0\}$ while $F=GF(q^{m})-\{0\}$,
then we can do the soft-decision rational interpolation based on $M(x,e)$.
\vspace{-0.2cm}
\begin{thm}
For $\theta\neq\infty$ and $Q(x,y)$ interpolates $(\alpha^{-i},p(\alpha^{-i},\alpha^{-j}))$
with multiplicity $M(\alpha^{-i},\alpha^{-j})$ for every $i,j\in\{1,2,3,\ldots,n\}$.
If $D=\deg_{1,L_{a}-L_{b}-1}(Q)$ and
\[
D+2(\deg(\Lambda)-L_{a})D_{Y}(Q)<\sum_{i\in I}M(\alpha^{-i},e_{i}),
\vspace{-0.5cm}
\] then $Q(x,g^{\prime}(x))=0$.\end{thm}
\begin{IEEEproof}
If $\theta\neq\infty$, $g^{\prime}(x)=\frac{uv^{\prime}-u^{\prime}v}{(u+\theta v)^{2}}$,
$\deg(uv^{\prime}-u^{\prime}v)\leq2\deg(\Lambda)-(L_{a}+L_{b})-1$
, $\deg((u+\theta v)^{2})\leq2(\deg(\Lambda)-L_{a})$; then the
maximum degree of numerator of $Q(x,g^{\prime}(x))$ is $D+2(\deg(\Lambda)-L_{a})D_{Y}(Q)$
like the bound in Theorem 2. For $i\in I$, $p(\alpha^{-i},e_{i})=g^{\prime}(\alpha^{-i})$,
so the numerator of $Q(x,g^{\prime}(x))$ passes $x=\alpha^{-i}$
with multiplicity $M(\alpha^{-i},e_{i})$. Once the degree of a polynomial
is less than the sum of its zero point multiplicity, it must be the
zero polynomial. So we prove the conclusion.\end{IEEEproof}
\vspace{-0.2cm}
\begin{cor}
For $\theta=\infty$, $Q(x,y)$ interpolates $(\alpha^{-i},p(\alpha^{-i},\alpha^{-j}))$
with multiplicity $M(\alpha^{-i},\alpha^{-j})$ for every $i,j\in\{1,2,3,\ldots,n\}$.
If $D=\deg_{1,L_{b}-L_{a}-1}(Q)$, and
\vspace{-0.2cm}
\[
D+2(\deg(\Lambda)-L_{b})D_{Y}(Q)<\sum_{i\in I}M(\alpha^{-i},e_{i})
\vspace{-0.4cm}
\]then $Q(x,g^{\prime}(x))=0$.
\end{cor}
During the interpolation to get $Q(x,y)$ , we have to solve
\vspace{-0.2cm}
\begin{equation}
C=\sum_{i=1}^{n}\sum_{j=1}^{n}\frac{M(\alpha^{-i},\alpha^{-j})(M(\alpha^{-i},\alpha^{-j})+1)}{2}\label{eq:C_count}
\end{equation}
equations in total. Then we can have the following conclusion.
\begin{thm}
For $\deg(\Lambda)\geq L_{b}$, if
\vspace{-0.2cm}
\begin{equation}
\frac{C+1}{D_{Y}+1}+D_{Y}(2\deg(\Lambda)-\frac{3L_{a}+L_{b}+1}{2})\leq\sum_{i\in I}M(\alpha^{-i},e_{i})\label{eq:AA}
\end{equation}
\vspace{-0.2cm}
and
\vspace{-0.2cm}
\[
(L_{b}+1-L_{a})D_{Y}(D_{Y}+1)/2\leq C+1
\] then there exist a non-zero polynomial $Q(x,y)$ with $D_{Y}(Q)\leq D_{Y}$, that $Q(x,y)$ interpolates the points $(\alpha^{-i},p(\alpha^{-i},\alpha^{-j}))$
with multiplicity $M(\alpha^{-i},\alpha^{-j})$ for $i,j\in[1,n]\cap Z$ and $Q(x,g^{\prime}(x))=0$.\end{thm}
\begin{IEEEproof}
If $\theta\neq\infty$, suppose $\deg_{1,L_{b}+1-L_{a}}(Q)=D$
and $D\geq0$, then $Q(x,y)$ can have $(D+1)(D_{Y}(Q)+1)+(L_{b}+1-L_{a})D_{Y}(Q)(D_{Y}(Q)+1)/2$ terms.
So there exists non-zero polynomial $Q(x,y)$ interpolate the points above with $D_{Y}(Q)\leq D_{Y}$ if
\[
(D+1)(D_{Y}+1)+(L_{b}+1-L_{a})D_{Y}(D_{Y}+1)/2\geq C+1
\]
Meanwhile, 
\[
\min(D)\leq \frac{C+1}{D_{Y}+1} - \frac{(L_{b}+1-L_{a})D_{Y}}{2}
\]

When
\vspace{-0.2cm}
\begin{equation}
D+2(\deg(\Lambda)-L_{a})D_{Y}+1\leq\sum_{i\in I}M(\alpha^{-i},e_{i})\label{eq:CC}
\vspace{-0.2cm}
\end{equation}, then $Q(x,g^{\prime}(x))=0$. So we complete the proof because (4) is stronger than \eqref{eq:CC} if $Q= \underset{Q}{\operatorname{argmin}}(D)$.

If $\theta=\infty$, then
\[
\frac{C+1}{D_{Y}+1}+D_{Y}(2\deg(\Lambda)-\frac{L_{a}+3L_{b}+1}{2})\leq\sum_{i\in I}M(\alpha^{-i},e_{i})\label{eq:AA}
\]
because $L_{a}\leq L_{b}\leq \deg(\Lambda)$. So the conclusion above is true when
$\theta=\infty$ as well.
\end{IEEEproof}

$\underset{Q}{\operatorname{argmin}}(D)$ will just have redundant
$y$ factor (or $D<0$) with unlimited or too large $D_{Y}$ setting, which is a corollary from the
negative weighted degree interpolation algorithm in \cite{Me}. So in the theorem we add an additional upper bound restrict for $D_{Y}$, in practical interpolation.

\subsection{Improved Interpolation Result and Corresponding Codewords Reconstruction
Algorithm in $GF(2^{m})$}

Usually, we only use RS code in $GF(2^{m})$ in practice. In this
field, the derivation of a polynomial must be a squared polynomial,
and every element must have a unique square root. So we do not need
to construct interpolation points for $g^{\prime}(x)$ but its square
root. We skip the proofs of the following conclusions because they
are almost the same as the previous two, except some degree arguments
change.
\begin{thm}
In $GF(2^{m})$, suppose $Q(x,y)$ interpolates $(\alpha^{-i},\sqrt{p(\alpha^{-i},\alpha^{-j})})$
with multiplicity $M(\alpha^{-i},\alpha^{-j})$ for every $i,j\in\{1,2,3,\ldots,n\}$.
If $\theta\neq\infty$, $D=\deg_{1,0.5(L_{a}-L_{b}-1)}(Q)$ and
\vspace{-0.3cm}
\[
D+(\deg(\Lambda)-L_{a})D_{Y}(Q)<\sum_{i\in I}M(\alpha^{-i},e_{i}),
\vspace{-0.2cm}\]
 then $Q(x,\sqrt{g^{\prime}(x)})=Q(x,\frac{\sqrt{uv^{\prime}-u^{\prime}v}}{u+\theta v})=0$;
or else if $\theta=\infty$, $D=\deg_{1,0.5(L_{b}-L_{a}-1)}(Q)$ and
\[
D+(\deg(\Lambda)-L_{b})D_{Y}(Q)<\sum_{i\in I}M(\alpha^{-i},e_{i}),
\vspace{-0.1cm}
\]
 then $Q(x,\sqrt{g^{\prime}(x)})=Q(x,\frac{\sqrt{-uv^{\prime}+u^{\prime}v}}{v})=0$.
\end{thm}

\vspace{-0.1cm}
\begin{thm}
In $GF(2^{m})$, for $\deg(\Lambda)\geq L_{b}$, if
\begin{equation}
\frac{C+1}{D_{Y}+1}+D_{Y}(\deg(\Lambda)-\frac{3L_{a}+L_{b}+1}{4})\leq\sum_{i\in I}M(\alpha^{-i},e_{i})\label{eq:codeword_condition}
\end{equation}

and \vspace{-0.3cm}
\begin{equation}
(L_{b}+1-L_{a})D_{Y}(D_{Y}+1)/4\leq C+1\label{eq:GMD},
\end{equation}
 then there exist polynomial $Q(x,y)$ with $D_{Y}(Q)$ not greater
than $D_{Y}$, that it interpolates the points $(\alpha^{-i},\sqrt{p(\alpha^{-i},\alpha^{-j})})$
with multiplicity $M(\alpha^{-i},\alpha^{-j})$ for every $i,j\in\{1,2,3,\ldots,n\}$ and $Q(x,\sqrt{g^{\prime}(x)})=0$.
\end{thm}

Because the denominator of $\sqrt{g^{\prime}(x)}$ may has factor
$x$, we should not directly factorize $Q(x,y)$ to get $\sqrt{g^{\prime}(x)}$.
If $\theta\neq\infty$ and we are concerned to $\Lambda$ that $\deg(\Lambda)\leq\rho$
and $\rho\geq b$ in practice, the worst case is $u+\theta v=x^{\rho-L_{a}}$.
So we should factorize the polynomial $Q(x,x^{-(\rho-L_{a})}y)$ to get $x^{\rho-L_{a}}\sqrt{g^{\prime}(x)}$
instead.

Once we get $g^{\prime}(x)$ , we can not recover $g(x)$ from $g^{\prime}(x)$
directly in finite field. But we can use two ``curves'' to get the
``intersection point''. Suppose $\theta_{1}\neq\infty$. For the
RS code in $GF(2^{m})$, we can construct the interpolation points
for both $\theta=\theta_{1}$ and $\theta=\theta_{2}$ to do interpolation
twice and get the two interpolation result $Q_{1}(x,y)$, $Q_{2}(x,y)$.
As \eqref{eq:codeword_condition} is not related to the value of $\theta$,
when its precondition is satisfied, we can get both $\sqrt{g_{1}^{\prime}}=\frac{\sqrt{uv^{\prime}-u^{\prime}v}}{(u+\theta_{1}v)}$
from factorization of $Q_{1}(x,y)$ and $\sqrt{g_{2}^{\prime}}=\frac{\sqrt{uv^{\prime}-u^{\prime}v}}{(u+\theta_{2}v)}$
if $\theta_{2}\neq\infty$ or else $\sqrt{g_{2}^{\prime}}=\frac{\sqrt{-uv^{\prime}+u^{\prime}v}}{v}$
from $Q_{2}(x,y)$ . If $\sqrt{g_{1}^{\prime}}/\sqrt{g_{2}^{\prime}}$
is known, we can get $\frac{v}{u+\theta_{1}v}$. We will introduce
two factorization methods to get $\sqrt{g_{1}^{\prime}}/\sqrt{g_{2}^{\prime}}$.

The first one is algebraic. $Q_{1}(x,y)$ and $Q_{2}(x,y)$ can be
viewed as two univariate polynomials in polynomial ring for variable
$y$, with coefficients of elements in univariate rational function
field for variable $x$. $Q_{1}(x,zy)$ and $Q_{2}(x,y)$ have the
common factor $y-\sqrt{g_{2}^{\prime}}$ in the ring if $z=\sqrt{g_{1}^{\prime}}/\sqrt{g_{2}^{\prime}}$.
We can use Euclidean algorithm from $Q_{1}(x,zy)$ and $Q_{2}(x,y)$
to get $R(x,z)$ after elimination variable $y$ ($R(x,z)$ is the
\textbf{resultant} of $Q_{1}(x,zy)$ and $Q_{2}(x,y)$ for variable
$y$). By the property of Euclidean domain, $R(x,(\sqrt{g_{1}^{\prime}}/\sqrt{g_{2}^{\prime}})$)
also has factor $y-\sqrt{g_{2}^{\prime}}$. In another word, $R(x,\sqrt{g_{1}^{\prime}}/\sqrt{g_{2}^{\prime}})=0$
(this is the property of resultant). So we can factorize $R(x,z)$
to get the root $\sqrt{g_{1}^{\prime}}/\sqrt{g_{2}^{\prime}}$. The
$z$-degree of $R(x,z)$ is no more than $D_{Y}^{2}/2$.

The other algorithm needs to do rational factorization twice. We can
factorize both $Q_{1}(x,y)$ and $Q_{2}(x,y)$ for variable $y$.
Then $\sqrt{g_{1}^{\prime}}/\sqrt{g_{2}^{\prime}}$ must be the division
result that one root of $Q_{2}(x,y)$ divided by one root of $Q_{1}(x,y)$.
So we can test all the division result of the pair combination of their roots(no more than
$D_{Y}^{2}/2$ times). Though this algorithm may output the same result
as the previous after we list all the probable division result, the
$y$-degree of $Q_{1}(x,y)$ and $Q_{2}(x,y)$ is not greater than
$D_{Y}\ll D_{Y}^{2}/2$ and the complexity of factorization reduces significantly.

Once we get $\sqrt{g_{1}^{\prime}}/\sqrt{g_{2}^{\prime}}$, we can
get $G(x)=(\theta_{2}-\theta_{1})^{-1}(\frac{\sqrt{g_{1}^{\prime}}}{\sqrt{g_{2}^{\prime}}}-1)$
if $\theta_{2}\neq\infty$ or else $G(x)=\frac{\sqrt{g_{1}^{\prime}}}{\sqrt{g_{2}^{\prime}}}$.
Then we need to check whether $G(\alpha^{-i})=h(\alpha^{-i})$ for
$i=1,2,\ldots,n$ to construct the error location set $I$. If the
evaluation of $\frac{\sqrt{g_{1}^{\prime}}}{\sqrt{g_{2}^{\prime}}}$
is $\frac{0}{0}$ type, we should remove the common factor in the
numerator and denominator. After we get $I$, if $\left|I\right|\leq n-k$,
we can interpolate any correct $k$ positions to get the codeword
polynomial; else for $i\in I$, we can use \eqref{eq:error value}
to evaluate the error value $e_{i}$ , then check whether it is a
valid error pattern.

For the RS code in a field with character beyond two, however, we
can only get $g_{1}^{\prime}/g_{2}^{\prime}$ after factorization
of the two interpolation result. To detect the error location, we
have to factorize the denominator and numerator of $g_{1}^{\prime}/g_{2}^{\prime}$
to get $\sqrt{g_{1}^{\prime}}/\sqrt{g_{2}^{\prime}}$ . This is because
we can not compare the evaluation of $g_{1}^{\prime}/g_{2}^{\prime}=(\frac{u+\theta_{2}v}{u+\theta_{1}v})^{2}$
and $(\frac{b-\theta_{2}a}{b-\theta_{1}a})^{2}$ if $\theta_{2}\neq\infty$,
because different values in this field may have same square result.

\vspace{-0.2cm}

\section{Y Degree, Multiplicity Assignment in $GF(2^{m})$}

In K{\"o}tter-Vardy algorithm, a multiplicity assignment algorithm
\cite[Algorithm A]{KV2003} is given to ensure the performance of
the decoding algorithm. We will prove that multiplicity assignment
algorithm can also be used in our decoding algorithm. For practical
reason, we just discuss it in $GF(2^{m})$.

As $\deg(\Lambda)=\left|I\right|$ , \eqref{eq:codeword_condition} is equal to
we have
\vspace{-0.2cm}
\begin{equation}
(\frac{C+1}{D_{Y}+1}-\sum_{i\in I}(M(\alpha^{-i},e_{i})-D_{Y}))D_{Y}^{-1}\leq\frac{3L_{a}+L_{b}+1}{4}\label{eq:key-event}
\end{equation}

Larger $D_{Y}$ setting  makes left side of  \eqref{eq:key-event} smaller under fixed $M, I$ and $C$. So we can set
{\vspace{-0.1cm}
\[
D_{Y}=\left\lfloor \sqrt{\frac{4(C+1)}{L_{b}+1-L_{a}}+\frac{1}{4}}-\frac{1}{2}\right\rfloor
\]
in practice during the interpolation in the theorem 9. We may notice if $M(\alpha^{-i},\alpha^{-j}) > D_{Y}$, then event \eqref{eq:codeword_condition}
will have larger probability to be true by $e_{i}=\alpha^{-j}$ rather than $e_{i}=0$. The count of such $(i,j)$ pair will be less than $2(L_{b}+1-L_{a})^{-1}$ as $(L_{b}+1-L_{a})(D_{Y}+2)(D_{Y}+1)>4(C+1)$ and \eqref{eq:C_count}. So the count is zero, or possible one unless $L_{b}=L_{a}$ and we can safely ignore it, like under assumption $\deg(\Lambda)\geq L_{b}$ we first flipping the hard decision $\hat{z}_{i}$ to
$\hat{z}_{i}+\alpha^{-j}$, the most possible error value before further decoding.

Suppose the memoryless channel outputs symbols $Z$ and the receiver
receives a vector $z$ in $Z^{n}$. We denote hard-decision of $z$
is $\hat{z}$. Error pattern vector $e$ is the subtraction of $\hat{z}$
and the sending codeword. We use a $n\times n$ matrix $\Pi$ to store
the probability $\Pi(\alpha^{-i},e_{i})=\Pr(\hat{z}_{i}-e_{i}|z_{i})$ for the $n$ non-zero error values at $n$
positions, $e_{i}\neq 0$ and $\Pr(\hat{z}_{i}-e_{i}|z_{i})$ denotes the posteriori probability of event of sending  
$\hat{z}_{i}-e_{i}$ at location $i$ after we received symbol $z_{i}$ at location $i$.
The multiplicity assignment algorithm is to find a multiplicity assignment
function $M(x,e)$ according to matrix $\Pi$. We denote the output
of function $M(x,e)$ also by a $n\times n$ integer matrix $M$ for $e\neq0$
and we define $M(x,0)=D_{Y}$. Then we can define the random variable
for the left side of \eqref{eq:key-event}
\vspace{-0.2cm}
\[
W(M)=(\frac{C+1}{D_{Y}+1}-\sum_{i=1}^{n}(M(\alpha^{-i},e_{i})-D_{Y}))D_{Y}^{-1}
\]

We can use Markov inequality to maximize the lower bound of probability
of \eqref{eq:key-event}. In another word, we want to minimize the
expectation of $W(M)$ like K{\"o}tter-Vardy's method under fixed
$C$ and $D_{Y}$.

Using the theory model in \cite{chernoff_multiplicity,KV2003}, we
research all the coset codes of the RS code on average instead to
simplify the model. Then error values $e_{i}$ ($i=1,2,\ldots,n$)
can be viewed as random variables with prior probability in uniform distribution in $GF(2^{m})$ (including zero). By
the linearity of expectation, the expectation of $W(M)$, $\mathbf{E}(W(M))$,
is
\vspace{-0.2cm}
\[
((\frac{C+1}{D_{Y}+1}-\sum_{i=1}^{n}\sum_{j=1}^{n}\Pi(\alpha^{-i},\alpha^{-j})(M(\alpha^{-i},\alpha^{-j})-D_{Y}))D_{Y}^{-1}
\]
\vspace{-0.4cm}

So we just need to get $M$ to maximize the inner product of matrix $M$
and $\Pi$, $\left\langle M,\Pi\right\rangle $. In another word,
we can reuse the multiplicity assignment algorithm\cite[Algorithm A]{KV2003}
to get the best $M$ and minimize the expectation of $W(M)$ from $\Pi$.

Using the K{\"o}tter-Vardy's multiplicity assignment algorithm, when
$C\rightarrow\infty$, then $M\rightarrow s\Pi$ (still $M(x,0)=D_{Y}$)  from \cite{chernoff_multiplicity} if we enable the elements
of $M$ has real number and $s$ is a number
related to $C$. Then $D_{Y}\rightarrow2\sqrt{\frac{C}{L_{b}+1-L_{a}}}$,
$s\rightarrow\sqrt{\frac{2C}{\left\langle \Pi,\Pi\right\rangle }}$
because of \eqref{eq:C_count}. So when $C\rightarrow\infty$, we can conclude from \eqref{eq:key-event}
asymptotically that our algebraic soft-decision decoding algorithm
can output an error vector $e$ if but not only if
\vspace{-0.2cm}
\begin{equation}
\deg(\Lambda)-L_{a} \leq\sqrt{\frac{L_{b}+1-L_{a}}{2\left\langle \Pi,\Pi\right\rangle }}\sum_{i\in I}\Pi(\alpha^{-i},e_{i})\label{eq:final_result}
\end{equation}

\vspace{-0.2cm}
It's not easy to give a fair compare between \eqref{eq:final_result}
and \cite[Theorem 12]{KV2003}, either measured by performance or complexity. But when $\Pr(\hat{z}_{i}-e_{i}|z_{i})\ll \Pr(\hat{z}_{i}|z_{i})$, K{\"o}tter-Vardy's
algorithm usually fails to get a codeword to correct the error at location $i$, while our algorithm \textbf{may} correct it  because $\Pr(\hat{z}_{i}|z_{i})$ and posteriori probability of sending the hard decision at other locations are not considered in interpolation, though we still use all the information from $z_{i}$ because $\sum_{k=0}^{n}\Pr(\hat{z}_{i}-k|z_{i})=1$. So at least, this algorithm can be considered as an alternative method
for RS decoding when other decoding algorithms fail.

\IEEEspecialpapernotice{\vspace{-18.5cm}
}

{\vspace{-0.3cm}
\section{CONCLUSIONS }

This paper briefly introduces a new algebraic soft-decision decoding
algorithm for RS code based on rational interpolation and factorization,
answering the open problem in \cite{Wu2008}. But there are several
important work left. We need more detailed performance analyze for
the presented multiplicity assignment algorithm according to Markov
inequality. We need to find some better multiplicity assignment algorithms,
such as Chernoff bound type algorithm like \cite{chernoff_multiplicity}
and types method like \cite{types_multiplicity} for our decoding
algorithm. The factorization needs better algorithm to reduce the
complexity. An open question is left: can we find the re-encoding
algorithm like \cite{re-encoding} to reduce the complexity of rational
interpolation?

\IEEEspecialpapernotice{\vspace{-2cm}
}
\vspace{-0.2cm}

\bibliographystyle{IEEEtran}
\bibliography{algebraic_coding}

\begin{flushleft}

\par\end{flushleft}

\end{document}